\documentclass[final,reprint,twocolumn,floatfix,notitlepage,prb,superscriptaddress]{revtex4-1}
\usepackage{graphicx,amssymb,soul,color,amsmath,bm}
\usepackage{epstopdf,hyperref}
\usepackage{braket,dsfont}
\usepackage[mathcal]{euscript}
\usepackage{comment}

\hypersetup{colorlinks=true,citecolor=blue,linkcolor=magenta}

\sethlcolor{yellow}
\allowdisplaybreaks

\begin{document}

\title{Many-body effects in porphyrin-like transition metal complexes embedded in graphene}

\author{Andrew Allerdt}
\affiliation{Department of Physics, Northeastern University, Boston, Massachusetts 02115, USA}
\author{Hasnain Hafiz}
\affiliation{Department of Physics, Northeastern University, Boston, Massachusetts 02115, USA}
\affiliation{Department of Mechanical Engineering, Carnegie Mellon University, Pittsburgh, PA 15213, USA}
\author{Bernardo Barbiellini}
\affiliation{School of Engineering Science, Lappeenranta University of Technology, Lappeenranta, Finland}
\affiliation{Department of Physics, Northeastern University, Boston, Massachusetts 02115, USA}
\author{Arun Bansil}
\affiliation{Department of Physics, Northeastern University, Boston, Massachusetts 02115, USA}
\author{Adrian E. Feiguin}
\affiliation{Department of Physics, Northeastern University, Boston, Massachusetts 02115, USA}

\begin{abstract}
We introduce a new computational method to study porphyrin-like transition metal complexes, bridging density functional theory and exact many-body techniques, such as the density matrix renormalization group (DMRG). We first derive a multi-orbital Anderson impurity Hamiltonian starting from first principles considerations that qualitatively reproduce GGA+U results when ignoring inter-orbital Coulomb repulsion $U'$ and Hund exchange $J$. An exact canonical transformation is used to reduce the dimensionality of the problem and make it amenable to DMRG calculations, including all many-body terms (both intra, and inter-orbital), which are treated in a numerically exact way. We apply this technique to FeN$_4$ centers in graphene and show that the inclusion of these terms has dramatic effects: as the iron orbitals become single occupied due to the Coulomb repulsion, the inter-orbital interaction further reduces the occupation yielding a non-monotonic behavior of the magnetic moment as a function of the interactions, with maximum polarization only in a small window at intermediate values of the parameters. Furthermore, $U'$ changes the relative position of the peaks in the density of states, particularly on the iron $d_{z^2}$ orbital, which is expected to greatly affect the binding of ligands.
\end{abstract}
\maketitle

\section{Introduction}

Porphyrins and metalloporphyrins attract a great deal of interest due to their crucial role in biological processes such as respiration and photosynthesis. 
These, and similar molecules such as phtalocyanine, are polyaromatic complexes that can accommodate a range of atoms in their center, giving them different magnetic and optical properties\cite{RawatRef6,RawatRef7,RawatRef8,Asano2016a}.
Thanks to their versatility, they have found a range of exciting applications in spintronics\cite{Wende2007,Bogani2008,Heutz2013,Bernien2009,Bhandary2011,Zeng2014,Mittra.Second_sphere,Sahoo.Hydrogen,Li2018,Rubio-Verdu2018}, optoelectronics\cite{Lee2006,Sergeyev2007}, solar cells\cite{Yella2011,RawatRef1,RawatRef3,RawatRef4,RawatRef5,Higashino2016}, and as building blocks of magnetic materials\cite{RawatRef9,RawatRef10,Shimizu2003,Yamashita2011,Saha-Dasgupta2014,Rawat2015,Lepper2017} or highly tunable qubits for quantum computing applications\cite{Urtizberea2018}. 
Experiments and subsequent theoretical treatments have also shown important correlation physics, such as spin and orbital variants of the Kondo effect in phtalocyanine (FePc) molecules deposited on the (111) surface of noble metal \cite{Iancu2006,Tiago2009,Minamitani2012,Lobos2014,Huang2014,Wang2015, Fernandez2015,Fernandez2018}. %A recent proposal suggests it might be possible to realize topologically protected excitonic states in porphyrin thin films\cite{Yuen-Zhou2014}.
In an exciting development, porphyrin-like centers can be embedded in graphene and carbon nanotubes to be used for oxygen reduction catalysis\cite{Zhang2009,Lee2011,Chung2013,Zhu2013,Orellana2013,Jia2015,Tylus2016,Chen2017,Aoyama2018}, and it is reasonable to think of a number of potential applications mimicking Nature but in a large scale and with increased tunability. In addition, transition metals can be a source of magnetism\cite{Lee2012}, providing another knob for realizing unconventional functionality.
%Work towards spintronic devices using magnetic ordered porphyrins has been shown to be promising\cite{Wende2007}. They show the magnetic moment of iron porphyrin molecules on ferromagnetic nickel and cobalt films will order ferromagnetically and can be controllably rotated in and out of the plane.
%More recently, it has been demonstrated that switching of the spin state can be achieved in complexes by altering the ligand bonds with the introducing hydrogen bonding\cite{Mittra.Second_sphere,Sahoo.Hydrogen}.

Despite their apparent simplicity of graphene and the transition metal complex, understanding their combined electronic structure remains a challenge. The center typically consists of a transition metal atom with a incomplete $d$ shell that give rise to confinement induced correlations and magnetism. %In addition, their small size represent a difficult problem for conventional electronic structure methods.
Traditionally, the study and simulation of a transition metal complex, iron porphyrins, or heme-like molecules use density functional theory (DFT) \cite{Groot1998,Johansson2004,Scherlis.Simulation_heme,Kramm2012,Wu2013,Kattel2012,Kattel2013,Berryman2015,Jia2015}, quantum Monte Carlo\cite{Koseki2008,Aspuru-Guzik2004}, coupled-cluster\cite{Johansson2004}, molecular dynamics\cite{Rovira2000} or configuration interaction techniques\cite{Labute.Strong_electron_corr}. Although these types of problems have been analyzed for the past couple of decades, a unique novel approach will be taken here. Using DFT calculations as a benchmark and comparison,  the combination of the density matrix renormalization group (DMRG)\cite{White1992,White1993,DMRGbook,Schollwock2005,Feiguin2013a} along with a unitary transformation will be employed to account for the many-body physics in a numerically exact way. Due to the large number of different geometries, transition metals, and axial ligands, a versatile method to account for strong correlations will be useful as each configuration will exhibit different properties and possible applications. 

%GGA+U studies of iron porphyrin-type molecules have revealed that interaction effects on the iron play an important role in determining the ground state magnetic moment\cite{Pooja.GGAU_modeling,Scherlis.Simulation_heme,Weber5790}. While GGA calculations with $U=0$ captures the magnetic ground state quite well, $U$ is needed for describing the energy splittings in the spectrum\cite{Kattel2012}. Predicting a physical effective onsite Coulomb repulsion of $U\approx4 ~\mathrm{eV}$, and a Hund interaction of $J\approx 1\mathrm{eV}$, the spin-state of the molecule is seen to change with $U$ with a high spin state for large values of the interaction strength, as expected. However, this is ignoring the inter-orbital Coulombic effects which will be shown to be crucial. 

%In DFT, the effective one-electron potential that enters the Kohn-Sham equation contains a many body term called exchange-correlation potential which is unknown. 
Finding accurate approximations to the exchange-correlation functional represents the central problem of DFT. The simplest exchange-correlation potential is the local density approximation (LDA), which is assumed to be a function of the local electron density only \cite{Jones1989}. This LDA energy functional has permitted the calculation of the ground-state properties of the 3d magnetic metals, including magnetic moments and Fermi surfaces, and the results are generally in good agreement with experiments \cite{Mackintosh1980}. Despite these success, LDA is not good enough to describe the phase diagrams of magnetic materials. In particular, it does not reproduce correctly the lowest-energy crystal structure of pure iron. LDA calculations predict that the the non-magnetic, face-centered cubic structure of iron has a lower energy than the ferromagnetic, body-centered cubic structure. The Generalized Gradient Approximation (GGA) \cite{Perdew1996} provides a simple but in principle more accurate step beyond LDA, which includes the effect of the density gradient in the exchange-correlation functional. Interestingly, the GGA correctly predicts the relative stability of the ferromagnetic phase of pure Fe, as well as giving a very good description of its ground state properties \cite{Barbiellini1990}. However, there are still remaining problems with the GGA. In fact, GGA+U studies of iron porphyrin-type molecules have revealed that interaction effects on the iron play an important role in determining the ground state magnetic moment\cite{Pooja.GGAU_modeling,Scherlis.Simulation_heme,Weber5790}.

\begin{figure}
\centering
\includegraphics[scale=0.56]{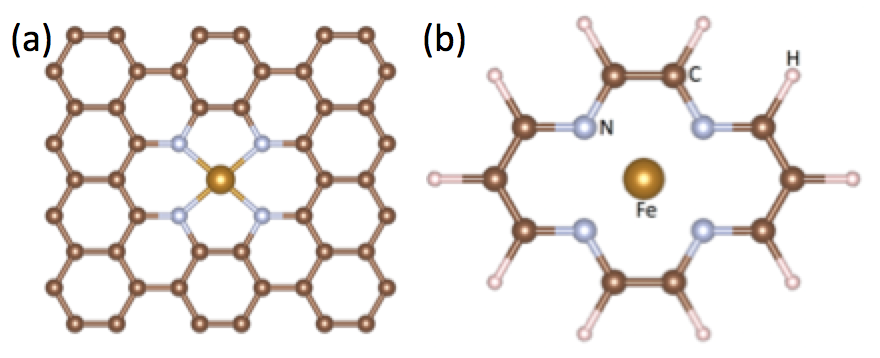}
\caption{(a) A FeN$_4$ complex embedded in a graphene sheet. (b) The $\mathrm{FeC}_{10}\mathrm{N}_4$ complex functionalized by 10 hydrogen atoms.}
\label{molecule}
\end{figure}
In this paper, we consider a FeN$_4$ center embedded in a graphene lattice. The fundamental bulding block will be considered as a FeN$_4$C$_{10}$H$_{10}$ complex (referred-to as $D1$ center in Ref.\onlinecite{Aoyama2018}). These centers have been experimentally prepared in graphene and carbon nanotubes \cite{Lee2011,Jia2015}. We adopt this configuration since it requires minimum structural modification to the graphene backbone and provides with the minimal unit that allows for an easy comparison between different numerical approaches. While GGA calculations with $U = 0$ capture the molecule's magnetic ground state quite well\cite{Kattel2012}, $U$ is needed for describing the energy splittings in the spectrum. Predicting a physical effective onsite Coulomb repulsion of $U \approx 4$ eV, and a Hund interaction of $J \approx 1$eV, the spin-state of the molecule is seen to change with $U$ with a high spin state for large values of the interaction strength, as expected. However, this is ignoring the inter-orbital Coulomb effects which will be shown to be crucial. We discuss the derivation of the five-orbital Kanamori-Anderson effective Hamiltonian in Sec.~\ref{model}, and the method used to solve it in Sec.~\ref{method}. The solution for the molecule and graphene and the corresponding phase diagrams are described in Sec.~\ref{results}. We finally conclude with a summary and discussion.

\section{Model Hamiltonian}\label{model}

In order to study the role of many-body correlation effects in transition metal complexes we will derive a simplified model that accounts for the most relevant features that play a role in determining the electronic structure and occupation of the transition metal atom. For this purpose, correlation effects are ignored in the carbon and nitrogen atoms.
In this work we construct the organic backbone of the molecule under consideration using an LCAO (or tight-binding) Hamiltonian, while using exact numerical methods to account for all the many-body physics introduced by the central atom, which we model as a multi-orbital Anderson-like impurity.
Generalized Anderson impurity models have already been applied to porphyrin-like molecules\cite{Tiago2009,Manoranjan2012,Thomas2013,Lobos2014,Fernandez2015,Fernandez2018} and have been able to predict potential energy surfaces and electronic coupling factors\cite{Labute.Strong_electron_corr,Labute.Anderson_imp} of different transition metal complexes. The advantage of solving a model Hamiltonian are numerous, but most remarkably: (i) we can account for all the many body correlation effects in a numerically exact way, and (ii) we can easily scale it to multiple impurities and more complex geometries. 

We present our approach by starting from a graphene sheet, a two-dimensional arrangement of carbon atoms on a honeycomb lattice. 
A transition-metal center is created by removing six carbon atoms and replacing them with four nitrogens and a single iron atom, as shown in Fig.~\ref{molecule}(a).
For simplicity we assume the complex is planar and has point group symmetry $D_{2h}$. To determine the active orbitals, we first start by considering the carbon atoms and we build a model for graphene. The carbon atoms are connected by $\sigma$-bonds of hybrid $sp^2$ orbitals, formed from linear combinations of its $2p_x$, $2p_y$, and $2s$ orbitals, which are responsible for its structural properties. Weaker than the $\sigma$-bonds are $\pi$-bonds, formed by the remaining $p_z$ orbitals that are mainly responsible for the electronic properties. At its edges, the graphene sheet is functionalized by hydrogen atoms bonded with each dangling $sp^2$ orbital of the carbons. 

\begin{figure}
\centering
\includegraphics[width=0.45\textwidth]{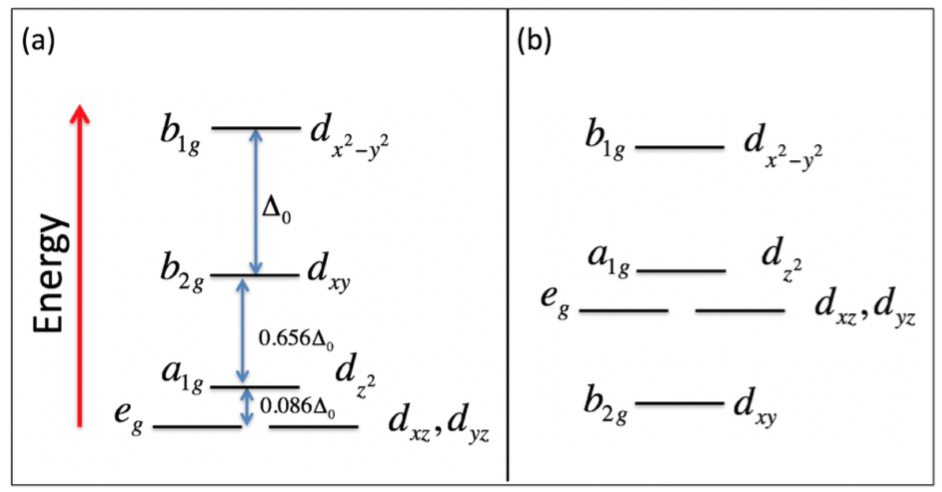}
\caption{(a) Splitting of $d$ energy levels in $D_{4h}$ symmetry. $\Delta_0$ is the splitting in an octahedral field. (b) Arrangement of energy levels used throughout this work. Note that in Ref.~\cite{Scherlis.Simulation_heme}, this corresponds to five-coordinated iron. The difference between the $\pi$ and $z^2$ orbitals is small and has been reversed to match the occupation of the orbitals according to DFT calculations.}
\label{splittings}
\end{figure}

To approach these bonds in a tight-binding manner, the values of the hoppings will be approximated to that of bulk graphene. Therefore, modeling the $p_z$ orbitals require only a simple nearest neighbor hybridization $t$, resulting in the well known two-band model of graphene. Graphene's $\sigma$ bands require a more sophisticated approach. For details of the derivation of the tight-binding Hamiltonian for the $\sigma$ bands, we refer to our Appendix A.
As an approximation, the nitrogen atoms are treated on equal footing as the carbons, meaning the C-C hopping is equal to the N-C hopping. The nitrogens however will have remaining orbital pointing towards the iron.

Focusing exclusively on the iron and nitrogen center, the complex is approximately square planar and has $D_{4h}$ symmetry, which will be useful in determining the bonding orbitals and crystal field splitting. Utilizing only the $3d$-orbitals of the iron, the $d_{x^2-y^2}$ will form a $\sigma$ bond with the dangling $sp^2$ orbital from the nitrogen, while $\pi$-bonds will from from the iron $d_{xz}$, $d_{yz}$ and nitrogen $2p_z$ orbitals (we consider the $x$ and $y$ axes poingint along the lines connecting the Fe and the N atoms).
Group theory predicts the level splitting of the five $d$ orbitals. The five dimensional $(l=2)$ irreducible representation of $K$ (the continuous rotation group), becomes reduced to 4 irreducible representations when the symmetry is decreased to $D_{4h}$.
% as,
%\begin{equation}
%D^{(l=2)}=a_{1g} + b_{1g} + b_{2g} + e_g \quad .
%\end{equation}
%The first three being 1-dimensional and the last 2-dimensional. Each representation corresponds to a different energy level as stated by the Wigner theorem. The representation $e_g$ corresponds to degenerate basis functions $d_{xz}$ and $d_{yz}$. The representations $a_{1g}$, $b_{2g}$, and $b_{1g}$ are formed by $d_{z^2}$, $d_{xy}$, $d_{x^2-y^2}$  respectively. Determining the absolute value of these splittings and levels is a much more complicated matter. 
Accounting for only electrostatic effects, the crystal field splitting is predicted to be that seen in Fig.~\ref{splittings}(a). However, this does not capture higher order effects or Jahn-Teller distortions\cite{JahnTeller}. For a more accurate description, the levels in a graphitic structure are arranged as in Refs.~\cite{Scherlis.Simulation_heme, Aoyama2018}, as shown in Fig.~\ref{splittings}(b). As we describe below, the positions of the energy levels are adjusted to approximately match the occupations of the $3d$ orbitals of the DFT data at $U=4.0$eV. They are then held constant throughout all other calculations.

Inside the iron, we are able to include Coulomb and Hund interactions. The most general form of the Hamiltonian can be written as\cite{Dagotto.Colossal,deMedici2017}:
\begin{eqnarray}
\label{interactions}
H_{int}=\sum_{{m_1,m_2,m'_1,m'_2}\atop{\sigma_1,\sigma_2,\sigma'_1,\sigma'_2}}\bra{m_1m_2\sigma_1\sigma_2}V\ket{m'_1m'_2\sigma'_1\sigma'_2} \times \\ \nonumber 
\times d^\dagger_{m_1\sigma_1}d^\dagger_{m_2\sigma_2}d_{m'_2\sigma'_2}d_{m'_1\sigma'_1}
\end{eqnarray}
In the above expression, $m$ labels the $d$-orbitals and $V\left(\bf{r}\right)$ is the screened Coulomb potential. This can be simplified to 4 matrix elements known as Kanamori parameters\cite{Kanamori.electron_correlation}: the intra-band Coulomb interaction $U$, the inter-band Coulomb interaction $U'$, the inter-band exchange interaction $J$, and the pair hopping amplitude $J'$. It can be shown\cite{Kotliar.Interplay_mott_ferro} that $J=J'$. This is due to the symmetry of the orbitals and the fact that all coefficients are just integrals of the Coulomb term over the radial part of the wave functions. In order to ensure rotational invariance in orbital space, the condition $U=U'+2J$ must be satisfied\cite{Dagotto.Colossal}. Effects of the crystal field on the Coulomb interactions have been ignored and are assumed to be relatively small. The interaction part of the Hamiltonian now takes the following form:
\begin{gather}
H_{int}=U\sum_{m}n_{\uparrow m}n_{\downarrow m} + \sum_{m> m'} \left( U'_{mm'}-\frac{J_{mm'}}{2}\right)n_{m}n_{m'} \nonumber \\
- 2\sum_{m> m'}J_{mm'}\vec{S}_m\vec{S}_{m'} - \sum_{m> m'}J_{mm'}d^\dagger_{\uparrow m}d^\dagger_{\downarrow m}d_{\uparrow m'}d_{\downarrow m'}
\end{gather}
The interactions $U$, $U'$, and $J$ can alternatively be expressed in terms of the so called Racah parameters $A$, $B$, and $C$, as shown in Table \ref{table:UJ}. The values of $B$ and $C$ are given in Ref.~\onlinecite{Weissbluth.hemoglobin} for Fe$^{3+}$ and Fe$^{2+}$ and are restated in Table \ref{table:racah}.

\begin{table}

\centering
\begin{tabular}{|c | c||c | c|}
  \hline
  m & m$'$ & $U'$ & $J$ \\
  \hline
  xy, xz, yz & xy, xz, yz & A-2B+C & 3B+C\\
  \hline
  xz, yz & z$^2$ & A+2B+C & B+C\\
  \hline
  xz, yz & x$^2$-y$^2$ & A-2B+C & 3B+C\\
  \hline
  xy & z$^2$ & A-4B+C & 4B+C\\
  \hline
  xy & x$^2$-y$^2$ & A+4B+C & C\\
  \hline
  x$^2$-y$^2$, z$^2$ & x$^2$-y$^2$, z$^2$ & A-4B+C & 4B+C\\
  \hline
\end{tabular}
\caption{Values of $U'$ and $J$ for the different $3d$ orbitals considered.\cite{Dagotto.Colossal}}
\label{table:UJ}
\end{table}

\begin{table}
\centering
\begin{tabular}{|r|c|c|}
  \hline
  Ion & B$(eV)$ & C$(eV)$ \\
  \hline
  Fe$^{2+}$ & 0.114 & 0.501 \\
  \hline
  Fe$^{3+}$ & 0.126 & 0.595 \\ 
  \hline
\end{tabular}
\caption{Racah Parameters for iron in different ionic states.}
\label{table:racah}
\end{table}

In our calculations, we assume that there are approximately 6 electrons in the iron orbitals and therefore take values close to that of Fe$^{2+}$. This leaves only one free parameter for the interactions, i.e. $A$, or equivalently $U$, since all  $U$ and $U'$ have the same dependence on $A$. Note that $J$ is independent of $A$. In general, it is the competition between $U$ and $J$ that determines if the complex is high or low spin. 

In addition, spin orbit coupling can be introduced as it becomes important for heavier elements that could substitute the iron in similar molecules. 
Due to the fact that this interaction for iron is relatively small, it is ignored in our calculations.

%Interpreting the SOC term in the Schr\"odinger equation as an effective Zeeman term in the rest frame of the electron, and approximating the nucleus as a spherically symmetric potential, the spin orbit interaction can be written as
%\begin{equation}
%\label{soc}
%H_{SO}=\lambda \vec{L} \cdot \vec{S} \quad,
%\end{equation}
%where $\lambda$ contains the radial integration and is treated as a tight binding parameter. For a more detailed discussion and derivation, see Ref.~\cite{Konschuh.dissertation}. The onsite matrix elements $\bra{l m}H_{SO}\ket{l m'}$ for the $d$-orbitals ($l=2$) are listed as:
%\begin{gather}
%\label{so_matrix}
%H_{SO} =  \nonumber \\
%\bordermatrix{~ & d_{xy} & d_{x^2-y^2} & d_{xz}  & d_{yz} & d_{z^2}  \cr
%                  d_{xy} & 0 & 2is_z & -is_x & is_y & 0  \cr
%                  d_{x^2-y^2} & -2is_z & 0 & is_y & is_x & 0  \cr
%                  d_{xz} & is_x & -is_y & 0 & -is_z & i\sqrt{3}s_y  \cr
%                  d_{yz} & -is_y & is_x & is_z & 0 & -i\sqrt{3}s_x \cr
%                  d_{z^2} & 0 & 0 & -i\sqrt{3}s_y & i\sqrt{3}s_x & 0 \cr}
%\end{gather}

\section{Method}\label{method}

Once all the parameters for the model are obtained, the problem can be recast and solved using the DMRG method. In order to do this, we will map the problem onto an equivalent one-dimensional model by employing an exact canonical transformation, as presented by two of the authors in Refs.~\cite{Feiguin.Lanczos} and \cite{Allerdt.Kondo}, and reviewed in detail in Ref.~\onlinecite{Allerdt2019}. The general method will be outlined here. 

A conventional Hamiltonian for impurity problems will have the form
\begin{equation}
\label{hami1}
H=H_l + H_{imp} + V ~.
\end{equation}
Here, $H_l$ represents a single-particle tight-binding Hamiltonian for the lattice, which could include more than one band and be obtained from DFT simulations. $H_{imp}$ and $V$ describe the impurity and the coupling between impurity and lattice, respectively. Note this method is applicable regardless of the geometry or dimensionality of the non-interaction Hamiltonian. The central concept is to map $H_l$ onto an equivalent one-dimensional chain. First, for simplicity, let us consider a single impurity problem and one orbital per site. More general cases of multiple orbitals and impurities will be discussed in the next section. The first step is to define the ``seed'' state to perform a Lanczos recursion as
\begin{equation}
\label{seed}
\ket{\Psi_0}=c^\dagger_{r_0}\ket{0} ~,
\end{equation}
where $c^\dagger_{r_0}$ creates an electron at orbital $r_0$, and $\ket{0}$ is the vacuum state. 
For an Anderson-like impurity (such as the case here), the seed is chosen as the impurity orbital.
Next, the rest of the states are constructed with the following iterative procedure:
\begin{gather}
\ket{\Psi_1}=H_l\ket{\Psi_0}-a_0\ket{\Psi_0} \label{lanczos1} \\
\ket{\Psi_n+1}=H_l\ket{\Psi_n}-a_n\ket{\Psi_n}-b^2_n\ket{\Psi_{n-1}} \\
a_n=\frac{ \bra{\Psi_n}H_l\ket{\Psi_n} }{ \braket{\Psi_n|\Psi_n} } \quad
b^2_n=\frac{ \braket{\Psi_n|\Psi_n} }{\braket{\Psi_{n-1}|\Psi_{n-1}}} \quad .
\end{gather}
The equations for $a_n$ and $b_n$ are obtained by requiring the states to be orthogonal. Note, however, that at this stage the states are not normalized.

\begin{figure}
\centering
\frame{\includegraphics[scale=0.26]{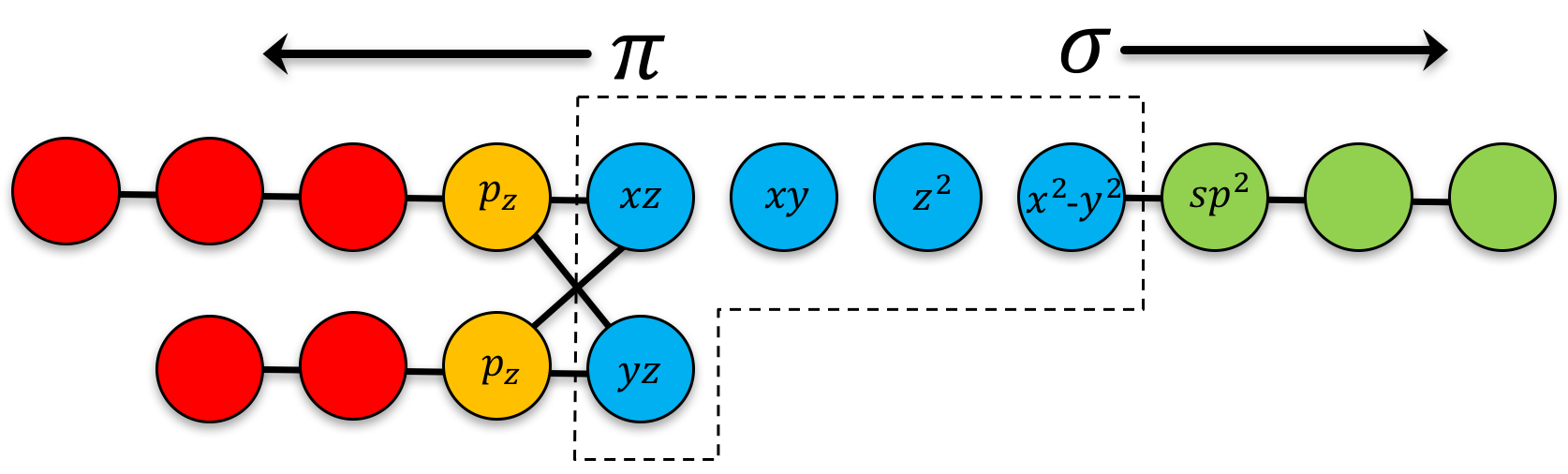}}
\caption{Chain geometry of FeC$_{10}$N$_4$ after the Lanczos mapping. Red sites correspond to carbon atoms. Orange sites correspond the seeds orbitals $\ket{\alpha_0}$, and $\ket{\beta_0}$ on the nitrogen atoms (see text). Blue sites represent the 5 $d$-orbitals of iron. The dotted line box represents where interactions are included, while black lines represent hoppings.  }
\label{chain_Fe}
\end{figure}

After this transformation, $H_l$ has a tri-diagonal form:
\begin{equation}
\label{tridiagonal}
H_l=\begin{pmatrix}
a_{0} & b_{1} & 0 & 0  \\
b_{1} & a_{1} & b_{2} & 0 \\
0  & b_{2}  & a_{2} & b_3  \\
0 & 0 & b_3 & \ddots
\end{pmatrix} ~,
\end{equation}
which corresponds to the geometry of a chain, a one-dimensional Huckel Hamiltonian.
Equivalently, in second quantization it reads
\begin{equation}
H_l=\sum_{i=0}^{L}a_i\tilde{n}_i +\sum_{i=0}^{L-1}b_{i+1}(\tilde{c}^\dagger_i\tilde{c}_{i+1} + h.c.) \quad ,
\end{equation}
where $\tilde{c}^\dagger_i$, $\tilde{c}_i$ are normalized creation and destruction operators respectively, $\tilde{n}_i=\tilde{c}^\dagger_i\tilde{c}_i$ is the particle number operator, and $L$ is the total length of the chain.
The diagonal $a_n$ terms are on-site potentials, while the $b_n$'s are the new hoppings along the chain. 
%While performing this iteration procedure in practice, numerical errors are introduced for large $L$ due to finite precision. There are two ways to reduce these errors: 1) Since $\braket{\Psi_n|\Psi_n}$ is growing with $n$, normalize $\Psi_{n-1}$ at each iteration. 2) To prevent loss of orthogonalization, orthogonalize each vector with all previous vectors at each iteration. 

This is indeed an exact canonical transformation. The remaining missing orbitals correspond to different symmetry sectors of the Hamiltonian and are completely decoupled from the impurity and can be safely ignored, which highlights the power of the change of basis. 
%Therefore, as mentioned, the impurity will only couple to this single site of the chain.

%At its core, the idea of the mapping is similar to that of NRG. The main difference being that in NRG, the mapping onto a chain is done in energy space. Here we have the advantage of obtaining, with high precision, spatial resolution of one or more impurity interactions. It also permits the study of any lattice structure, or Hamiltonian that is quadratic.  

Coming back to the transition metal complex, our present case of interest has two orbitals (iron $d_{xy}$ and $d_{yz}$) coupled to different sites (nitrogens) that will generate two orthogonal chains using a technique very similar to that described above but requiring two seeds for the $\pi$-bonding (nitrogen $p_z$ orbitals). Labeling them as $\ket{\alpha_0}$, and $\ket{\beta_0}$, they are chosen to be:
\begin{equation}
\label{seeds}
\ket{\alpha_0}=\frac{1}{2}
\begin{pmatrix}
  1\\
  1 \\
  1\\
  1 \\ \end{pmatrix} \quad
  \ket{\beta_0}=\frac{1}{2}
\begin{pmatrix}
  1\\
  -1 \\
  1\\
  -1 \\ \end{pmatrix}~,
\end{equation}
where the labeling corresponds to the nitrogen sites in Fig.~\ref{molecule}. {Notice that due to symmetry, only these two -out of four - wave-functions couple to $d$ orbitals of the transition metal. The remaining degrees of freedom live in an orthogonal Hilbert space that does not contribute to the physics, or chemistry, or the problem. } After the Lanczos iterations are carried out, two chains are generated and are represented by the red sites in Fig.~\ref{chain_Fe}. The coupling Hamiltonian between the nitrogen and iron then becomes,
\begin{equation}
\label{coupling}
H_{c}=-t'\{d_{xy}^\dagger(\ket{\alpha_0}+\ket{\beta_0}) + d_{yz}^\dagger(\ket{\alpha_0}-\ket{\beta_0}) \} \quad .
\end{equation}

For the $\sigma$-bands, the single seed mapping is used starting from the $d_{x^2-y^2}$ orbital. The hopping integrals between nitrogen and carbon's $2s$, $2p_x$, and $2p_y$ orbitals are obtained from the tight binding model described above. 
After this transformation, the green chain in Fig.~\ref{chain_Fe} is produced. { Once again we notice that, due to the symmetries of the problem and the resulting dimensional reduction, the total number of orbitals in the equivalent system is smaller than the original one. } 
The magnitude of the hoppings between iron and nitrogen's $sp^2$ orbitals can be used as a fitting parameter, while the $\pi$ coupling between the nitrogen and iron is estimated by comparing to DFT data to be $t'\approx 1.6eV$. To study the problem of the iron embedded in bulk graphene, the two sides of the chain in Fig.~\ref{chain_Fe} are just extended to the desired length as discussed in Ref.~\cite{Feiguin.Lanczos}.

\section{Results}\label{results}

Although the DFT method is surprisingly accurate, it is known to have difficulties to describe systems with strongly correlated electrons in open $d$ or $f$ shells. In particular, DFT might fail in predicting if the ground state has low, intermediate or high total spin polarization. This failure can be somehow mitigated if the spin contamination is allowed as in Ref.~\onlinecite{Kattel2012}.
In computational chemistry, spin contamination is the spurious mixing of different electronic spin-states. This effect can occur when the spatial parts of up and down spin-orbitals are permitted to differ, which is generally undesirable because the mixing of spin states does not occur if the system is isolated. However, it can sometimes alleviate the problem of predicting a wrong ground state, as mentioned above.
In addition, conventional DFT methods, such as LDA or GGA, fail to properly account for the Coulomb interactions between localized electrons. Calculations presented here were performed using GGA+U with the VASP package\cite{vasp1,vasp2}. One main source of error arises from the fact that the $U$ and $J$ terms in the Hamiltonian (or functional) are handled in a mean-field fashion, resulting in a single $U_{eff}=U-J$ parameter that usually is adjusted leaving results somewhat arbitrary.
For these reasons, in order to carry out a comparison with DFT+U calculations, we first ignore all many-body terms except for the intra-orbital Coulomb repulsion and the Hund coupling of the spins. In other words, $H_{int} = U\sum_{m}n_{\uparrow m}n_{\downarrow m} - 2\sum_{m\neq m'}J_{mm'}\vec{S}_m\vec{S}_{m'}$. 

\begin{figure}
\centering
\includegraphics[scale=0.35]{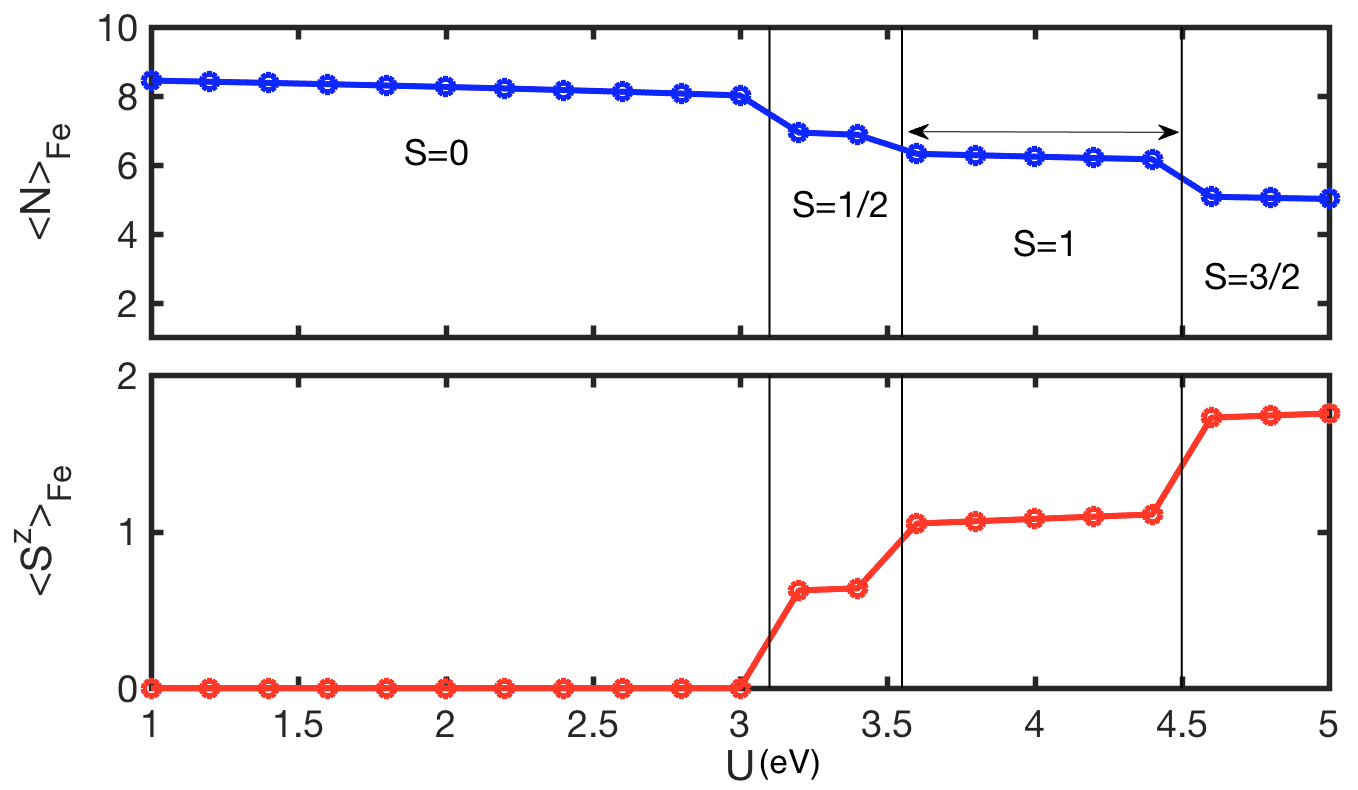}
\caption{Electronic occupation and magnetic moment of the iron atom with $U'=0$. Top panel shows the total occupation of the five iron orbitals. Bottom panel is the total value of $S^z$. The arrow indicates the $S=1$ phase consistent with experiment and DFT calculations.}
\label{noUp_diagram}
\end{figure}

To benchmark our approach, we begin by introducing the iron complex $\mathrm{FeC}_{10}\mathrm{N}_4$, depicted in Fig.~\ref{molecule}. 
{We have used the level splittings between the $d$ orbitals to match their electronic occupation with GGA+U results, using the physically relevant value of $U=4eV$, as shown on Table \ref{table:molecule}.} {Our formulation is $SU(2)$ invariant, meaning that any high spin ground state consists in reality of a $(2S+1)$-fold degenerate multiplet. }
The ground state occupation and magnetic moment of iron (total spin) as a function of $U$ can be seen in Fig.~\ref{noUp_diagram}. Calculations were done with varying the number of electrons and the value of the total spin $S$, from which we obtain the ground state by minimizing the energy. The spin remains zero until the Coulomb interaction reaches a value of $U\approx3.0eV$, where the effects of the repulsion become more relevant. From this point, the occupation of the iron levels will continue to decrease as the magnetic moment increases until the saturation value of $S=2$ ($\langle N \rangle =4$) is reached at large enough $U$. The large Coulomb repulsion prevents any orbital from being double occupied. In the range of $3.5 \leq U \leq 4.5$, the ground state has $S=1$ which is the physically interesting range. {Notice that, in principle, different Racah parameters should be used for each occupation of the iron atom. As described above, we have fixed them to those for Fe$^{2+}$}.

\begin{figure}
\centering
\includegraphics[scale=0.33]{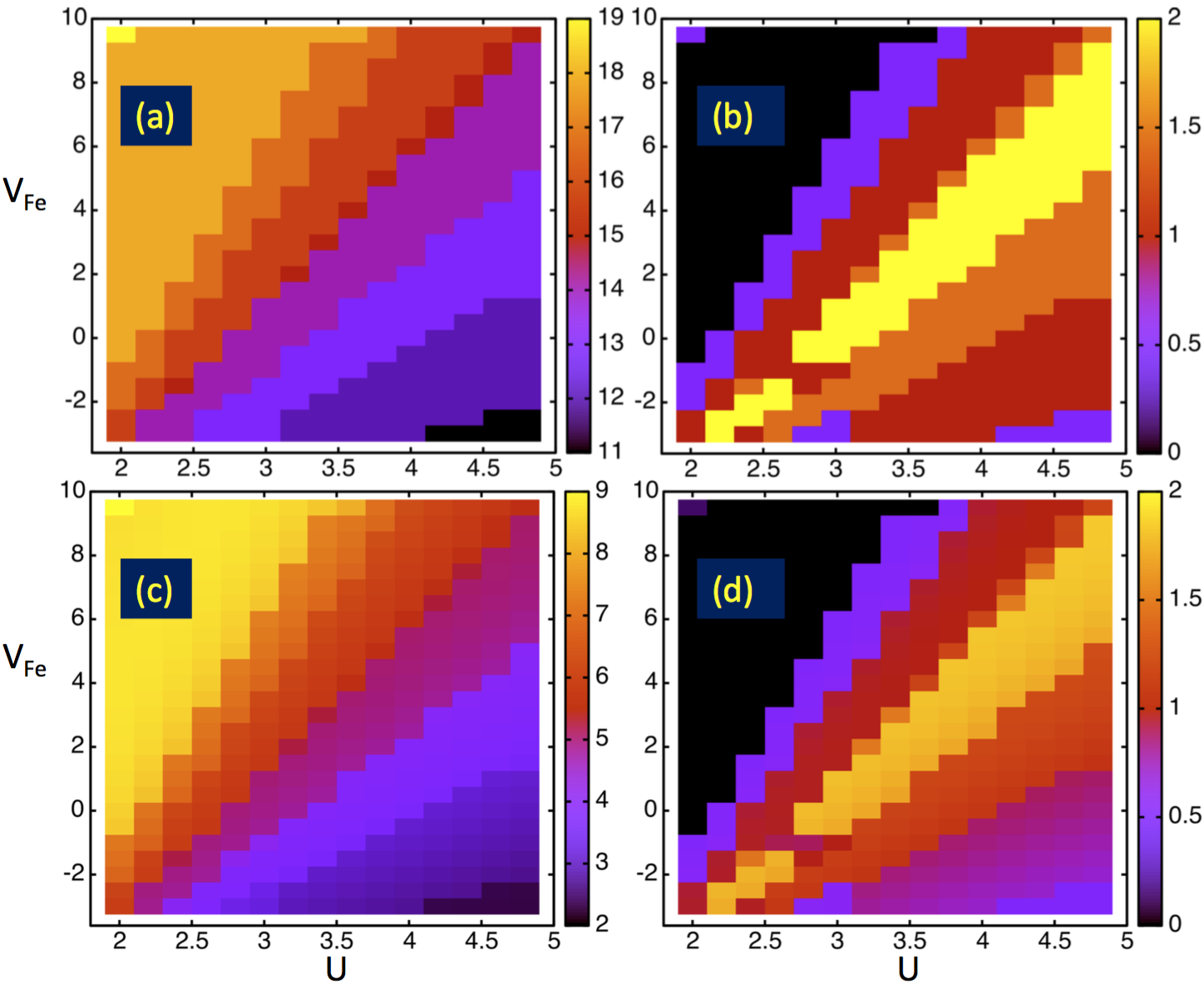}
\caption{Phase diagram of the full interaction Hamiltonian: (a) total occupation of the system; (b) total spin $S$; (c) occupation of the transition atom $\langle N \rangle_{Fe}$; (d) spin $\langle S^z \rangle_{Fe}$ of the transition atom in the maximally polarized state $S^z=S$. }
\label{phase_diagram}
\end{figure}

\begin{figure}
\centering
\includegraphics[scale=0.35]{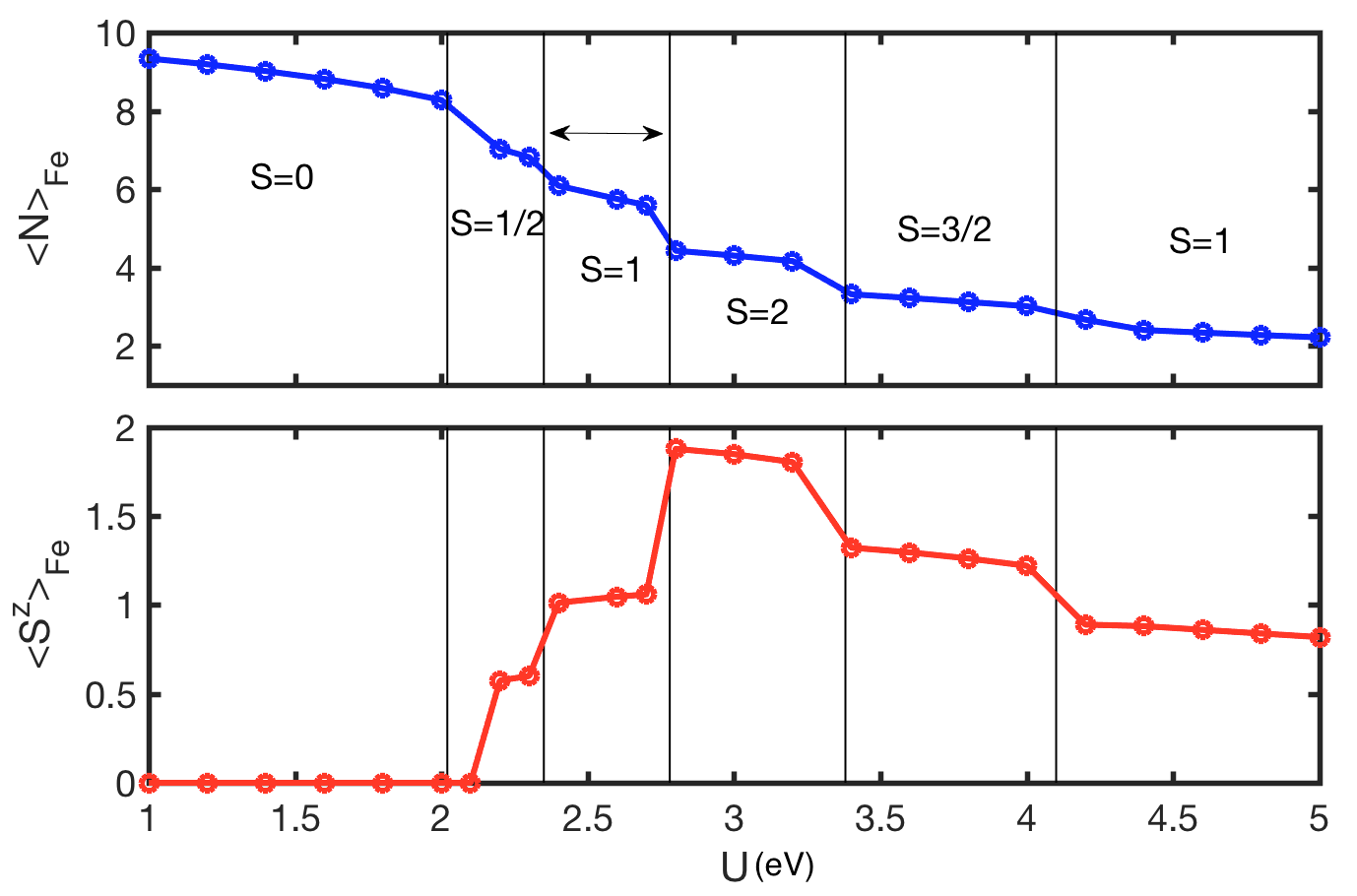}
\caption{Top panel shows the total occupation of the five iron orbitals as a function of $U$ and $V_{Fe}=0$ for the full interacting Hamiltonian. Bottom panel is the total value of $S^z$. The arrow indicates the physically interesting range corresponding to $S=1$.}
\label{Up_diagram}
\end{figure}

{
Including all terms in the interaction Hamiltonian changes the ground state drastically. To understand these effects we introduce an additional rigid shift in the position of the Fe energy levels $-V_{Fe}\sum_m{n_m}$. 
This potential is related to the (screened) interaction with the nucleus, and serves as a parameter to control the occupation of the levels and tune between different transition metals.  
In Fig.~\ref{phase_diagram} we plot the overall occupation of the molecule, the Fe atom, and the total spin $S$, as a function of $U$ and $V_{Fe}$. 
We find that the parameter regime of interest for Fe$^{2+}$ ($S=1$) resides in a narrow band of values, coinciding with the electronic configuration $(d^2_{xy}d^3_\pi d^2_z$), so-called $C_{231}$ in Ref.~\onlinecite{Bhandary2016}. Other bands in the figure correspond to a different ionic state of iron or a different atomic species (we point out again that different Racah parameters should be used in those cases). Within each region, the occupation of the different energy levels does not vary much. We focus on the parameter regime corresponding to $V_{Fe}=0$, and show the total charge and spin of the transition metal atom in Fig.~\ref{Up_diagram}. It is clear that these quantities depend strongly on $U$, creating a rich structure. The physically relevant region with $S=1$ and $\langle N\rangle\sim 6$ occupies a small range $2.3 < U < 2.8$. More importantly, increasing $U$ further will eventually plateau to a low spin state ($S=1/2$ or $S=0$), as opposed to reaching a high spin state as in the previous case without inter-Coulomb repulsion. The high spin states in this model are limited to a finite window of $U$. This is due to the fact that as $U$ is further increased, all orbitals are single occupied and the inter-orbital interactions start playing a dominant role.
}

It was mentioned earlier that the actual molecule is functionalized by hydrogen atoms bonded to the dangling $sp^2$ bond of the carbons. Calculations were done to compare results with and without taking these into account. The outcome is to slightly modify the occupation of the $d_{x^2-y^2}$ orbital, and does not change the overall physics. This effect is practically irrelevant if the $U'$ terms in the Hamiltonian are ignored. In all cases, these effects do not drastically affect the overall spin states of the iron. Therefore, only results without the hydrogen are shown.

%%

%%table
\begin{table}
\centering
\begin{tabular}{|c|c|c|c|c|c|c|c|c|}
 \hline
  & \multicolumn{2}{|c|}{DFT+U} & \multicolumn{2}{c|}{$U'=0$} & \multicolumn{2}{c|}{Full H} \\
 \hline
orbital & $\langle N \rangle$ & $\langle S^z \rangle$ & $\langle N \rangle$ & $\langle S^z \rangle$ & $\langle N \rangle$ & $\langle S^z \rangle$ \\
 \hline
$xy$ & 1.79 & 0.05 & 2.00 & 0.00 & 2.00 & 0.00 \\
 \hline
$z^2$ & 1.07 & 0.40 & 1.00 & 0.50 & 1.0 & 0.50 \\
 \hline
$x^2-y^2$ & 0.83 & 0.04 & 0.77 & 0.04 & 0.53 & 0.03 \\
 \hline
$\pi$ & 1.35 & 0.26 & 1.26 & 0.27 & 1.11 & 0.26 \\
 \hline
\end{tabular}
\caption{Occupation and magnetic moment of each iron orbital. The first three columns are results from DFT+U calculations. The next three are obtained from the method described in the text with $U'=0$, while the final three columns include all inter-orbital interactions.}
\label{table:molecule}
\end{table}
%%table

\begin{figure}
\centering
\includegraphics[scale=0.38]{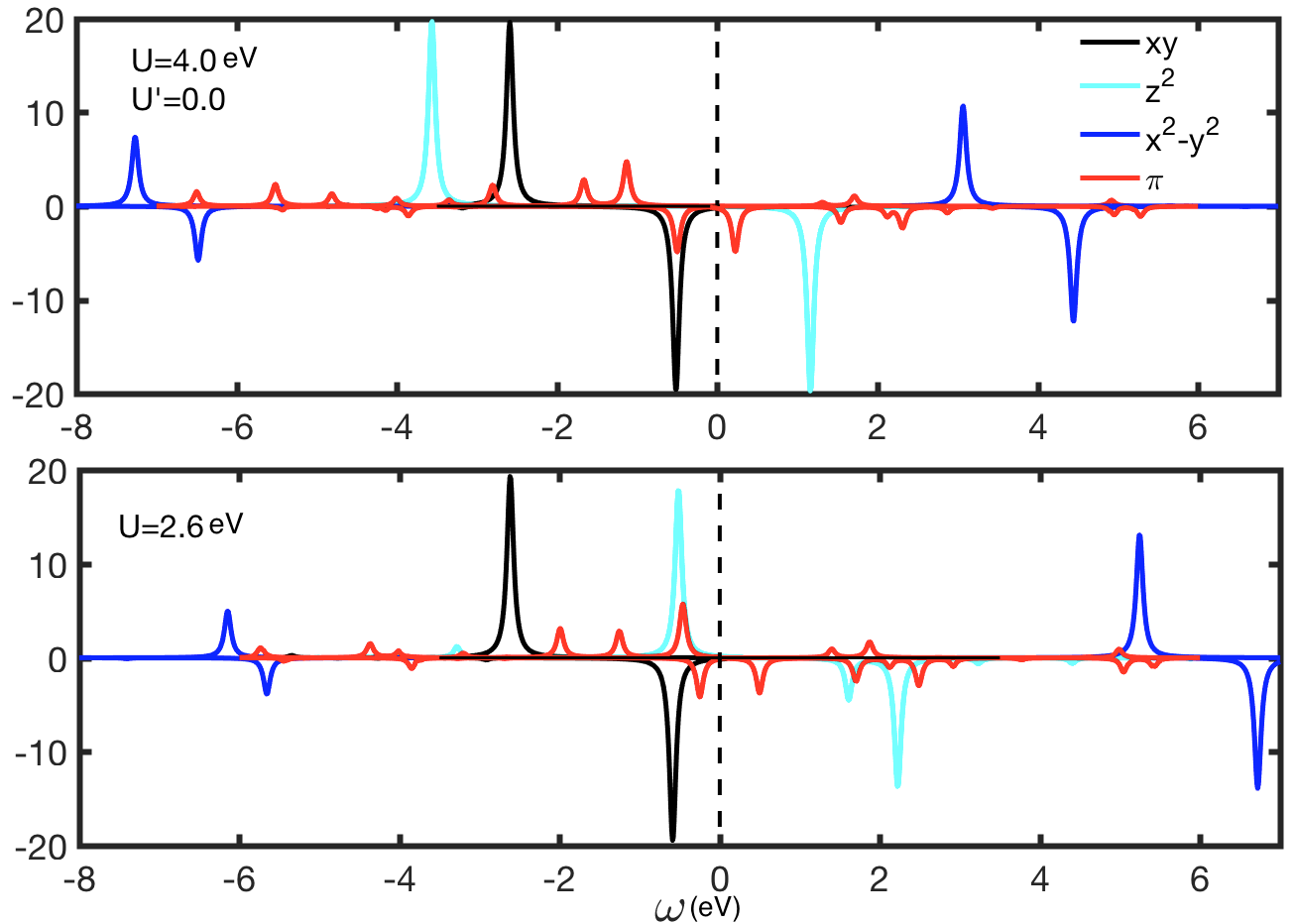}
\caption{Partial density of states for the $d$-orbitals of iron for spin $\uparrow$ (positive values) and $\downarrow$ (negative values). Top panel has $U'=0$ and the bottom panel has the full interaction Hamiltonian. The total number of electrons in both systems is $N_{tot}=16$.}
\label{dos_fe}
\end{figure}

\begin{figure}
\centering
\includegraphics[scale=0.38]{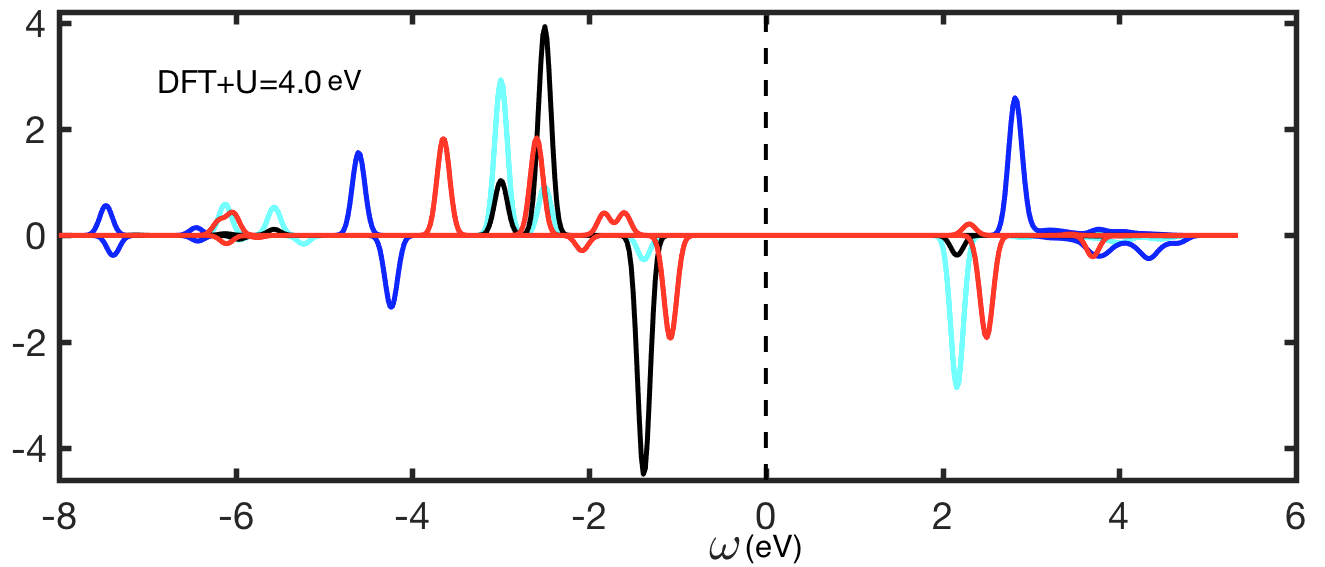}
\caption{Partial density of states for the $d$-orbitals of iron calculated with DFT+U with a GGA functional.}
\label{dft_fe}
\end{figure}

Figures \ref{dos_fe} and \ref{dft_fe} show the projected density of states for the five iron orbitals calculated with dynamical DMRG\cite{Kuehner1999,Jeckelmann2002} and DFT+U respectively. 
The top panel (without inter-orbital Coulomb interactions) shows close agreement with the DFT calculations. Parameters such as $U$, the energy levels of the $d$-orbitals, and the couplings were adjusted to agree with the occupation and magnetic moment given by DFT, since it is expected that DFT should capture most of the physics when inter-orbital interactions are ignored. The differences are possibly due to (i) approximations associated with DFT, (ii) the parameters utilized in our model, or (iii) the fact that we ignore Coulomb interactions between iron and nitrogen, and also in the rest of the system (even though they are somewhat accounted for by the effective hopping parameters).

Including the full interaction Hamiltonian modifies the LDOS, as one would expect. It is clear that the $U'$ terms shift the energies slightly upward, with the exception of the $d_{xy}$ orbital which is practically unchanged. Note however, that the value of $U$ is much reduced ($U=2.6$ compared to $U=4.0$), indicating the importance of these terms. Furthermore, the $d_{z^2}$ orbital is the one that is most greatly affected. One can see that this splitting between the peaks in Fig.~\ref{dos_fe} is controlled by $U$, while their relative positions with respect to other orbitals are dictated by $U'$ and $J$. This should have a great impact on binding of ligands which usually involve the $d_{z^2}$ orbital.

Extending the carbon atoms further by creating a ``bulk-like'' flake of graphene, causes the physics to change once again. Here, calculations are done including all interaction terms. For certain values of $U$, the spin actually increases compared to the molecule case. An intermediate $S=3/2$ phase appears around $U\approx2.6$ which was not previously present. This points toward the fact that the surrounding material plays a role in the physics on the transition metal atom: {a continuous density of states in bulk graphene, as opposed to just discrete ``delta''-like peaks, may allow for additional screening, in the same spirit as the Kondo effect}. 
%It could be possible to change the binding properties in these systems by changing the size and shape of the graphene flakes by electrostatic gating, for instance.

%\begin{table}
%\centering
%\begin{tabular}{|c|c|c|c|c|c|c|c|c|}
% \hline
% \multicolumn{3}{|c|}{$S=1$} & \multicolumn{3}{|c|}{$S=3/2$} & \multicolumn{3}{|c|}{$S=2$} \\
% \hline
%orbital & $\langle N \rangle$ & $\langle S^z \rangle$ & orbital & $\langle N \rangle$ & $\langle S^z \rangle$ & orbital & $\langle N \rangle$ & $\langle S^z \rangle$ \\
% \hline
%$xy$ & 2.00 & 0.00 & $xy$ & 2.00 & 0.00 & $xy$ & 1.00 & 0.50  \\
% \hline
%$z^2$ & 1.00 & 0.50 & $z^2$ & 1.00 & 0.50 & $z^2$ & 1.00 & 0.50 \\
% \hline
%$x^2-y^2$ & 0.53 & 0.03 & $x^2-y^2$ & 0.51 & 0.04 & $x^2-y^2$ & 0.50 & 0.04 \\
% \hline
%$\pi$ & 1.21 & 0.24 & $\pi$ & 1.01 & 0.36 & $\pi$ & 0.73 & 0.27 \\
% \hline
%\end{tabular}
%\caption{Occupation and magnetic moments of the $3d$ iron orbitals in a graphene flake. Values of $U$ are $U=2.5$, 2.6 and 3.0 corresponding to $S=1$, $3/2$, and 2, respectively.}
%\label{table:graphene}
%\end{table}

\begin{figure}
\centering
\includegraphics[scale=0.39]{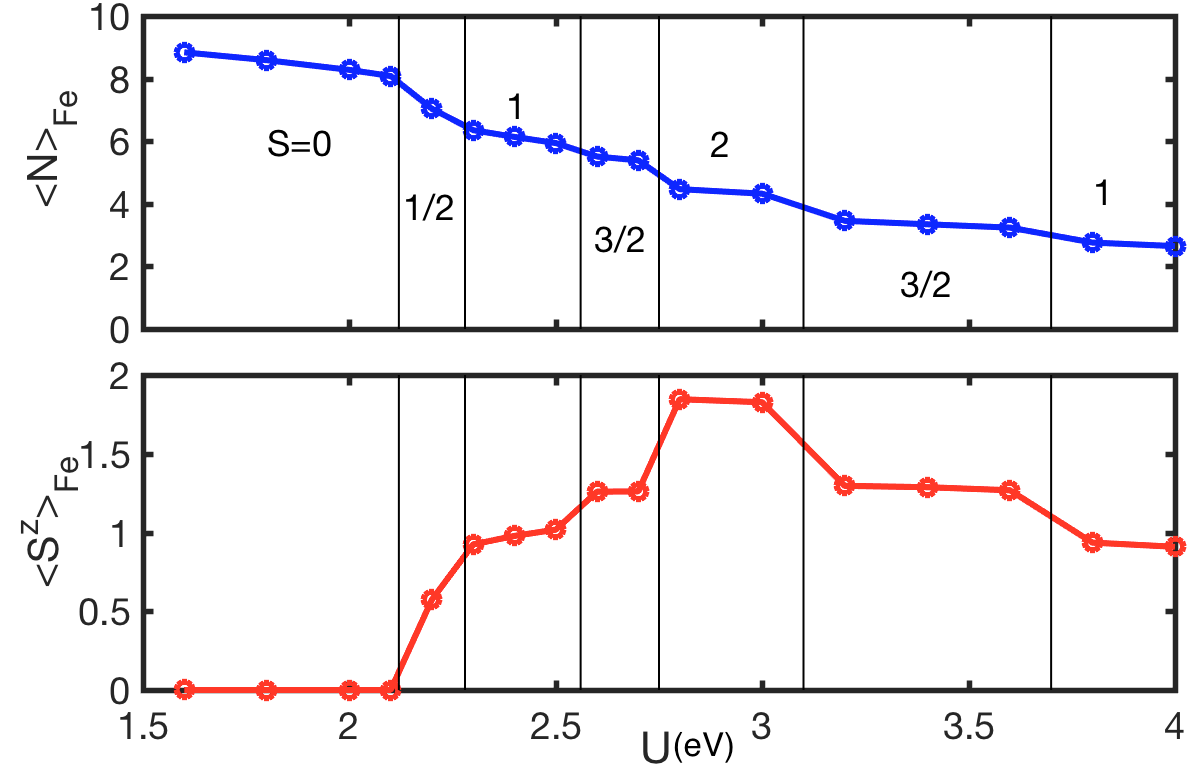}
\caption{Same as Fig.\ref{Up_diagram} but for a flake of graphene. Top pannel corresponds to the local occupation of the iron atom and bottom panel shows $\langle S^z\rangle_{Fe}$ as a function of $U$.}
\label{bulk_diagram}
\end{figure}

\section{Conclusions}\label{conclusions}
We have studied heme-like iron centers in graphene and the FeC$_{10}$N$_4$ molecule using an exact canonical transformation and the DMRG method. 
The DMRG technique has been used in quantum chemistry calculations as a solver for first principles Hamiltonians in the same spirit as configuration interaction\cite{Chan2011,Sharma2012,Sharma2014,Olivares-Amaya2015}. Our approach takes advantage of the weakly correlated nature of the carbon bond, which can be accounted for by DFT calculations, and recasts the problem onto an LCAO  model with an interacting transition metal center that is modeled as a 5-orbital Kanamori-Anderson impurity. This allows us to perform a unitary transformation that significantly simplifies the Hamiltonian, accounting for the most relevant degrees of freedom, and the multi-orbital nature of the problem.
 The resulting geometry consisting of one-dimensional chains coupled to the iron $d$-orbitals, makes it amenable to efficient DMRG calculations accounting for all many-body terms and treat them in a numerically exact way. We obtain the occupation and magnetic moment of the iron atom as a function of the Coulomb interaction $U$ and qualitatively recover DFT results when the inter-orbital repulsion is ignored. Upon including these terms, it is shown how crucial a role they play in the physics, by shifting the relative position of the peaks in the density of states. This is a dramatic effect that is expected to greatly affect binding of ligands.
%Using X-ray absorption, a hierarchy of mixed spin states\cite{Maltempo.spin_state} in iron porphyrins coined a ``magnetochemical series'' has been created\cite{Reed.Magnetochemical} in terms of ligand field strength. These mixed spin states create a continuum of magnetic moments on the iron atom, as opposed to pure integer or half-integer spin states. 
%The effects of correlated hybridization between the $d$-orbitals of the metal ions and ligands have also been investigated\cite{Cox.Correlated_hyrbid}, revealing important corrections in certain cases. These are ignored in the current work, but can easily be incorporated with the method presented.

In the future our technique can be combined with other quantum chemistry approaches such as CASPT2\cite{CASPT2,Yanai2017}, not only as a benchmark, but also to obtain realistic parameters to model the transition metal complex that can then be embedded in the bulk and mapped onto one dimensional chains.

%The newly found $S=3/2$ state in bulk graphene deserves further consideration, since it may indicate the presence of Kondo-like correlations, a quintessential many-body effect.

The method described in this work can be used to tackle related problems, as there are a variety of geometries and transition metals that could be studied. The inclusion of the effects of spin-orbit interactions and correlated hybridization\cite{Cox.Correlated_hyrbid} would be natural extensions. In addition, it is possible to consider two iron atoms in a sheet of graphene to investigate the emergence of any indirect magnetic exchange mediated by the conduction electrons. A powerful feature of our approach is that it can readily be extended to finite temperatures and adapted to study non-equilibrium phenomena such as transport and chemical reactions. %Arbitrary geometries could be considered, including gates to study transport, for instance. % If a reliable effective tight-binding model could be obtained, these types of molecules could also be placed on substrates. It is clear there are many possible systems that are able to be studied by following the prescription in this work.

\section*{Acknowledgements}
%We thank D. A. Scherlis for useful discussions. 
The work at Northeastern University was supported by the US Department of Energy (DOE), Office of Science, Basic Energy Sciences grant number DE-SC0019275.
%, and benefited from Northeastern University's Advanced Scientific Computation Center (ASCC) and the NERSC supercomputing center through DOE grant number DE-AC02-05CH11231.

\appendix
\section{Tight-binding model for graphene's $\sigma$-bands}
\begin{figure}
\centering
\includegraphics[scale=0.6]{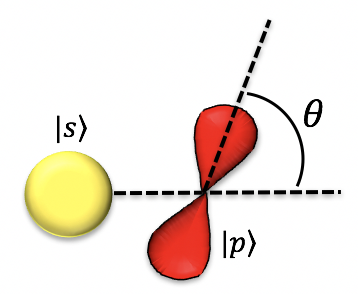}
\caption{Neighboring $s$ and $p$ orbitals showing the angle between the centers of the orbitals and the axis of the $p$ orbital.}
\label{sp_bond}
\end{figure}

Before discussing the electronic structure of the $\sigma$ bands, it is worthwhile to briefly present the general approach to determine the vanishing and non-vanishing matrix elements between orbitals. For only $s$ and $p$ orbitals, there are just four non-zero overlap integrals to consider: $ss\sigma$, $sp\sigma$, $pp\sigma$, and $pp\pi$. Due to the radial symmetry of the $s$-orbitals, the $ss\sigma$ bond has no angular dependence. $\sigma$ and $\pi$ bonds are classified by whether the interatomic separation and orbital axis are parallel or perpendicular. Generally, however, the $p$-orbitals can orient with any angle between them, as in Figure~\ref{sp_bond}. In this case, the orbitals are projected to their normal and parallel ( $\sigma$ and $\pi$ ) components. Therefore, the $p$-states can be decomposed as

$$
\ket{p}=\mathrm{cos}\theta \ket{p_{\sigma}} + \mathrm{sin}\theta \ket{p_{\pi}} \quad .
$$
Matrix elements between neighboring $s$ and $p$ states can then be written as 
$$
\bra{s}H\ket{p} = H_{sp\sigma}\mathrm{cos}\theta \quad,
$$
where the definition $\bra{s}H\ket{p_{\sigma}}=H_{sp\sigma}$ has been used, and $\bra{s}H\ket{p_{\pi}}=0$ by symmetry. The angle $\theta$ is defined in Figure~\ref{sp_bond}. Similarly, the matrix elements between $p$ states is seen to be
$$
\bra{p_1}H\ket{p_2} = H_{pp\sigma}\mathrm{cos}\theta_1 \mathrm{cos}\theta_2 + H_{pp\pi}\mathrm{sin}\theta_1 \mathrm{sin}\theta_2 \quad .
$$

In graphene, and in general, $sp^2$ orbitals are planar and form angles of $120\deg$. Since the unit cell of graphene has two atoms, and each atom contributes three $sp^2$ states, this method results in six $\sigma$-bands. Three of these lie below the Fermi level while three are above. Still following the recipe given by Ref.~\cite{graphene_book}, the $6\times6$ matrix has the following form:
\begin{equation}
\label{matrix6x6}
H = \bordermatrix{~ & 2s^a & 2p^a_x  & 2p^a_y  & 2s^b & 2p^b_x  & 2p^b_y \cr
                  2s^a & h_{11} & h_{12} & h_{13} & h_{14} & h_{15} & h_{16} \cr
                  2p^a_x & h_{21} & h_{22} & h_{23} & h_{24} & h_{25} & h_{26} \cr
                  2p^a_y & h_{31} & h_{32} & h_{33} & h_{34} & h_{35} & h_{36} \cr
                  2s^b & h_{41} & h_{42} & h_{43} & h_{44} & h_{45} & h_{46} \cr
                  2p^b_x & h_{51} & h_{52} & h_{53} & h_{54} & h_{55} & h_{56} \cr
                  2p^b_y & h_{61} & h_{62} & h_{63} & h_{64} & h_{65} & h_{66} \cr} \quad .
\end{equation}
The matrix elements are then given values as
\begin{gather}
h_{11}=h_{44}=\epsilon_s \\
h_{22}=h_{33}=h_{55}=h_{66}=\epsilon_p \\
h_{14}=H_{ss\sigma} \\
h_{15} = H_{sp\sigma}\mathrm{cos}\theta \\
h_{16} = H_{sp\sigma}\mathrm{sin}\theta \\
h_{25} = H_{pp\sigma} \mathrm{cos}^2\theta + H_{pp\pi} \mathrm{sin}^2\theta \\
h_{26} = (H_{pp\sigma} - H_{pp\pi})\mathrm{cos}\theta\mathrm{sin}\theta \\
h_{36} = H_{pp\sigma} \mathrm{sin}^2\theta + H_{pp\pi} \mathrm{cos}^2\theta \quad,
\end{gather}
with all remaining elements zero. The numerical values for the hoppings (in $eV$) are reported to be $\epsilon_s=-8.7$, $\epsilon_p=0$, $H_{ss\sigma}=-6.7$, $H_{sp\sigma}=5.5$, $H_{pp\sigma}=5.1$, and $H_{pp\pi}=-3.1$.

%\bibliography{Prelim,new,dmrg,ci,kondo,bernardo,porphyrins,vasp}

\end{document}